\documentclass[prl,reprint]{revtex4-1}

\usepackage{graphicx}
\usepackage{amsmath}

\DeclareMathOperator{\rect}{rect}
\DeclareMathOperator{\sinc}{sinc}

\begin{document}

\title{Reconciliation of generalized refraction with diffraction theory}

\author{St\'ephane Larouche}
\affiliation{Center for Metamaterials and Integrated Plasmonics, Department of Electrical and Computer Engineering,\\
Pratt School of Engineering, Duke University, Box 90291, Durham, North Carolina, 27708, United States of America}
\author{David R. Smith}
\affiliation{Center for Metamaterials and Integrated Plasmonics, Department of Electrical and Computer Engineering,\\
Pratt School of Engineering, Duke University, Box 90291, Durham, North Carolina, 27708, United States of America}

\date{\today}

\begin{abstract}

When an electromagnetic wave is obliquely incident on the interface between two homogeneous media with different refractive indices, the requirement of phase continuity across the interface generally leads to a shift in the trajectory of the wave. When a linearly position dependent phase shift is imposed at the interface, the resulting refraction may be described using a generalized version of Snell's law. In this Letter, we establish a formal equivalence between generalized refraction and blazed diffraction gratings, further discussing the relative merits of the two approaches.

\end{abstract}

\maketitle

The phase of an electromagnetic wave must be continuous across the homogeneous interface between two media. This constancy of phase underlies the phenomenon of wave refraction, in which the trajectory of a wave obliquely incident on the interface between two media with distinct refractive indices is altered as the wave crosses the interface. For plane waves, refraction is compactly and quantitatively expressed through Snell's law, which relates the incident and transmitted wave trajectories to the refractive indices of the two media. More simply stated, if the refractive indices of two media differ, the wavelengths of the waves propagating inside those media also differ, and the electromagnetic wave must bend at the interface to preserve the phase continuity (except for normal incidence). 

This simple picture breaks down if a phase shift is introduced at the interface. Such a phase shift can be created by a very thin layer through the use of metamaterial elements, subwavelength gratings, and many other similarly patterned structures. In general, the phase shift alters the continuity relation across the interface. If the phase shift is position independent, then the standard Snell's law remains valid. If the phase shift is position dependent, the refraction and reflection at the interface generally produce many plane waves propagating at various angles; for this case, refraction cannot be described by a single angle, and Snell's law is not applicable. 

For the special case of a phase shift that varies linearly as a function of the position along the interface, an incident plane wave remains intact and is bent by an amount proportional to the phase gradient. A thin layer with an artificially introduced linear gradient index can be used to steer a wave, as has been demonstrated in several experiments using metamaterial elements~\cite{Smith2005Gradient}. More recently, Yu {\em et al.} demonstrated beam steering at mid-infrared wavelengths using a single layer of lithographically patterned metamaterial elements to achieve the required phase gradient in the cross-polarization wave~\cite{Yu2011Light}. Ni {\em et al.} extended this work to telecommunication wavelengths~\cite{Ni2012Broadband}. For a purely linear gradient, the angles of the incident and transmitted waves can be related by a modified version of Snell's law that includes the phase gradient as an additional parameter.

\begin{figure}[b!]
	\includegraphics{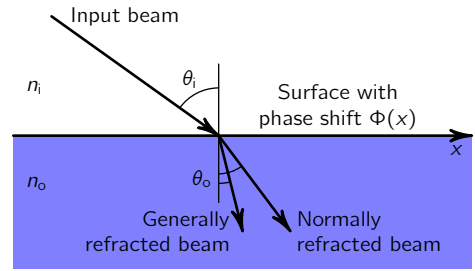}
	\caption{Schematic representation of general refraction. When a surface between two media introduces phase shift $\Phi(x)$ depending linearly on $x$, the beam is refracted at an angle different than that given by the standard Snell's law.}
	\label{FigGeneralizedRefraction}
\end{figure}

The configuration of interest, illustrated in Fig.~\ref{FigGeneralizedRefraction}, consists of two dielectric media separated by a thin gradient layer which is characterized by a phase shift $\Phi(x)$ that varies linearly as a function of a coordinate $x$ in the plane of the interface. A plane wave of wavelength $\lambda$ propagates in the plane containing the $x$ axis and the normal to the interface. For this case, the refracted wave is also a plane wave, and the propagation angles of the incident and refracted waves, $\theta_{\mathrm{i}}$ and $\theta_{\mathrm{o}}$, are related by
\begin{equation}
	n_{\mathrm{o}}\sin{\theta_{\mathrm{o}}}-n_{\mathrm{i}}\sin{\theta_{\mathrm{i}}} = \frac{\lambda}{2\pi}\frac{\mathrm{d}\Phi(x)}{\mathrm{d}x},
	\label{EqGeneralizedRefraction}
\end{equation}
where $n_{\mathrm{i}}$ and $n_{\mathrm{o}}$ are the refractive indices of the input and output media, respectively. A similar correction must be applied to calculate the angle of the reflected wave. While Eq.~\ref{EqGeneralizedRefraction} might be considered a generalized version of Snell's law, it is important to note that the equation actually encapsulates both refraction as well as wave propagation through an inhomogeneous medium (the gradient index layer). That a simple expression is realized is convenient, but rather specific to the profile considered. In general, neither refraction nor Snell's law can be strictly applied to an inhomogeneous interface.

A difficulty with a monotonically increasing or decreasing phase is the large total gradient required. Even with metamaterials, the available index differential is limited. The steering of a wave can still be achieved, however, by noting that, in the steady state regime, any phase shift $\Phi + k2\pi$, where $k$ is in integer, has the same effect as a phase shift $\Phi$. This degeneracy allows one to fold the linear $\Phi(x)$ into a triangular phase profile. A triangular phase profile was employed for the realization of the sample in Ref.~\onlinecite{Yu2011Light}, for example, as a practical means of achieving the linear profile.

Once periodicity has been introduced to the gradient profile, the possibility of diffraction cannot be ruled out. In fact, the triangular shape of the folded linear profile coincides with that of a blazed diffraction grating~\cite{Magnusson1978Diffraction}. A blazed grating is traditionally realized by varying the thickness of a grating following a saw-tooth profile in order to introduce a phase shift profile identical to that of the folded generalized refraction sample. The purpose of a blazed grating is to reduce or eliminate coupling of the incident wave to all save a single diffracted order. With a single plane wave produced in reflection or transmission, a blazed grating can alternatively be viewed as steering the incident beam. Blazed gratings can be realized by other means of introducing a saw-tooth phase profile, such as by varying the refractive index of a uniform thickness layer~\cite{Lalanne1998Blazed, Tsai2011Design}.

In the far field, the diffraction pattern is the Fourier transform of the complex transmission of the diffraction device. In the case of an infinitely periodic diffraction grating, the diffraction pattern is the product of a Dirac comb, and the Fourier transform of the motif of every period. The position of the diffraction peaks is given by the diffraction equation,
\begin{equation}
	n_{\mathrm{o}}\sin{\theta_{\mathrm{o}}}-n_{\mathrm{i}}\sin{\theta_{\mathrm{i}}} = m\frac{\lambda}{d},
	\label{EqDiffraction}
\end{equation}
where $d$ is the period of the grating and $m$, an integer, is the order of the diffraction peak. The Dirac comb can therefore be expressed as
\begin{equation}
	\sum_{m=-\infty}^{\infty}\delta\left(\alpha-\frac{m}{d}\right),
	\label{EqDiracComb}
\end{equation}
where $\delta()$ is the Dirac delta function and
\begin{equation}
	\alpha = \frac{n_{\mathrm{o}}\sin{\theta_{\mathrm{o}}}-n_{\mathrm{i}}\sin{\theta_{\mathrm{i}}}}{\lambda}.
	\label{Eqalpha}
\end{equation}

\begin{figure}[tb!]
	\includegraphics{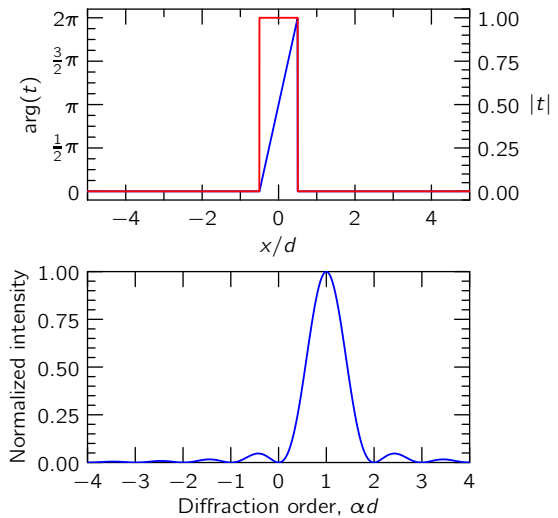}
	\caption{(top) Transmission amplitude (red) and phase (blue), and (bottom) diffraction pattern of the motif of a blaxed diffraction grating. The intensity has been normalized by $d^2$.}
	\label{FigMotif}
\end{figure}

If $\Phi(x)$ varies linearly by $2\pi$ over a distance $d$, then ${\mathrm{d}\Phi(x)}/{\mathrm{d}x} = 2\pi/d$ and Eq.~\ref{EqGeneralizedRefraction} predicts a generalized refraction angle equal to the first diffraction order ($m=1$). That case correspond to a motif that is a simply a linear phase change multiplied by a rectangular function 
\begin{equation}
	t_{\mathrm{motif}}(x) = \rect\left(\frac{x}{d}\right)\times\exp{{\mathrm i}2\pi\left(\frac{x}{d}-\frac{1}{2}\right)}.
\end{equation}
The diffraction pattern of this motif is
\begin{equation}
	{\cal F}\{t_{\mathrm{motif}}(x)\} = d\sinc\left(\alpha-\frac{1}{d}\right).
	\label{EqFTMotif}
\end{equation}
The motif and its diffraction pattern are shown in Fig.~\ref{FigMotif}. It should be noted that the diffraction pattern of the motif is 0 for all $\alpha = m/d$, except for $m=1$. Therefore, the diffraction pattern of the infinitely periodic blazed diffraction grating, which is the product of Eqs~\ref{EqDiracComb} and \ref{EqFTMotif}, is null everywhere except at $\alpha = 1/d$, exactly as predicted by generalized refraction.

\begin{figure}[tb!]
	\includegraphics{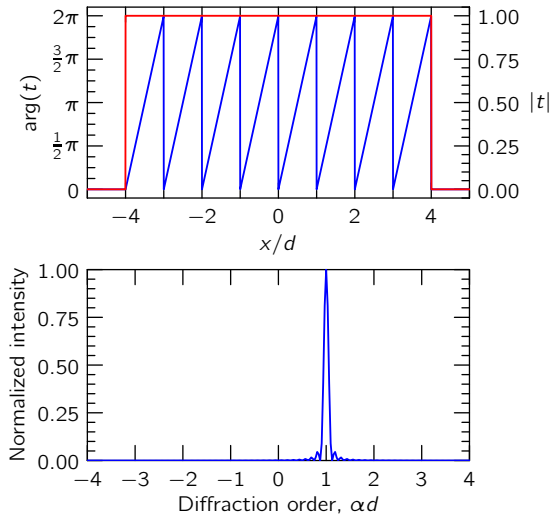}
	\caption{(top) Transmission amplitude (red) and phase (blue), and (bottom) diffraction pattern of a 8-period blazed grating. The intensity has been normalized by $(8d)^2$.}
	\label{FigBlazed}
\end{figure}

If the grating has a finite number of periods, every diffraction peak has a finite width and is a sinc function instead of being a Dirac delta function. Fig.~\ref{FigBlazed} shows the transmission and the diffraction pattern of a 8-period blazed diffraction grating.

\begin{figure}[tb!]
	\includegraphics{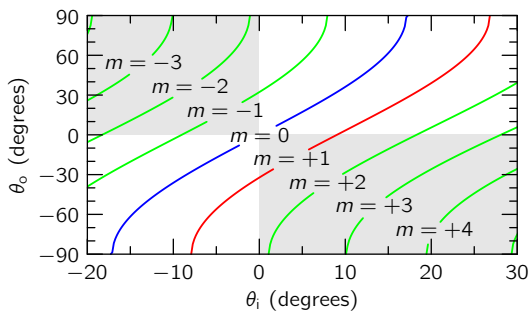}
	\caption{Diffraction angles for a wavelength of $8\,\mathrm{\mu{}m}$, a period is $15\,\mathrm{\mu{}m}$, a silicon input medium, and a air output medium. The $m=0$ (blue) line correspond to normal refraction while $m=1$ (red) corresponds to general refraction. Shaded regions correspond to {\em negative refraction}.}
	\label{FigAngles}
\end{figure}

Up to this point, the diffraction angles have been expressed by $\alpha$. Using appropriate values for $\theta_{\mathrm{i}}$, $n_{\mathrm{i}}$ and $n_{\mathrm{o}}$ one can easily calculate $\theta_{\mathrm{o}}$ using Eq.~\ref{Eqalpha}. Fig.~\ref{FigAngles} presents the diffraction angles in the conditions considered in Fig.~2 of Ref.~\onlinecite{Yu2011Light}. The first diffraction order correspond to the generally refracted beam. In certain conditions, the input and output angles are of different signs, which can be called negative refraction. However, it should be noted that this phenomenon is quite different from negative refraction occuring in negative refractive index metamaterials and cannot lead to perfect lenses~\cite{Pendry2000Negative}.

\begin{figure}[tb!]
	\includegraphics{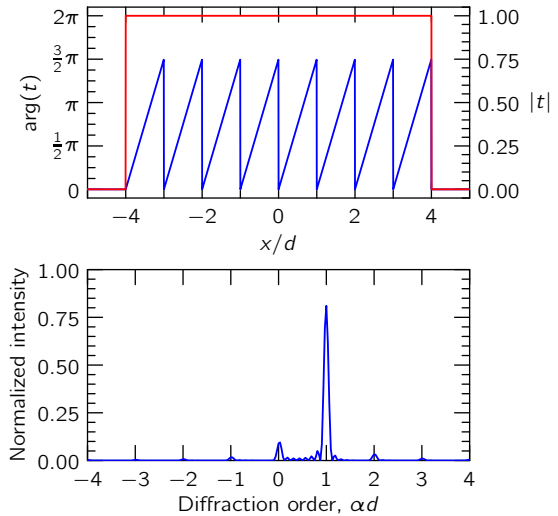}
	\caption{(top) Transmission amplitude (red) and phase (blue), and (bottom) diffraction pattern of a 8-period blazed grating with sub-optimal phase thickness. The intensity has been normalized by $(8d)^2$.}
	\label{FigBlazedNotEnoughPhase}
\end{figure}

If should be noted that diffraction theory can be applied even if the phase profile is not periodic. For the linear phase profile, the diffraction pattern is simply
\begin{equation}
	{\cal F}\left\{\exp{{\mathrm i}2\pi\frac{x}{d}}\right\} = \delta{}\left(\alpha-\frac{1}{d}\right),
\end{equation}
giving the exact same result as the infinite extent blazed diffraction grating. However, in addition to the practical reasons, we see an advantage in using a grating rather and a truly linear phase profile. In the former case, the diffraction angle is governed by the periodicity of the grating and is unaffected by errors in the phase profile. In the latter case, if $\Phi(x)$ does not have the correct slope, the position of the generally refracted beam is affected. For example, Fig.~\ref{FigBlazedNotEnoughPhase} shows a blazed grating with only $\frac{3}{2}\pi$ phase contrast. Some energy bleeds into other diffraction peaks, but the position of the first diffraction peak is unchanged.

Furthermore, we believe that the diffraction formalism has advantages over the general refraction formalism. For example, when realized with metamaterial elements, such as in Ref.~\onlinecite{Yu2011Light}, the phase profile needs to be discretized and only the diffraction formalism allows one to determine the effect of the discretization. The effect of many factors, such as the number of phase levels, absorption, and impedance mismatch, on the performance of metamaterial based blazed diffraction gratings has been studied recently~\cite{Smith2011Analysis}.

In conclusion, we believe that the generalized diffraction approach is valid, but works only in the simple case of a linear phase profile. The diffraction approach gives identical results for that case, but can also be applied to more complex phase profiles.

\section*{Acknowledgments}

This work was supported by a Multidisciplinary University Research Initiative (Grant No. W911NF-09-1-0539).

\bibliography{../../../../Literature/.bib/articles,../../../../Literature/.bib/books}

\begin{thebibliography}{8}%
\makeatletter
\providecommand \@ifxundefined [1]{%
 \@ifx{#1\undefined}
}%
\providecommand \@ifnum [1]{%
 \ifnum #1\expandafter \@firstoftwo
 \else \expandafter \@secondoftwo
 \fi
}%
\providecommand \@ifx [1]{%
 \ifx #1\expandafter \@firstoftwo
 \else \expandafter \@secondoftwo
 \fi
}%
\providecommand \natexlab [1]{#1}%
\providecommand \enquote  [1]{``#1''}%
\providecommand \bibnamefont  [1]{#1}%
\providecommand \bibfnamefont [1]{#1}%
\providecommand \citenamefont [1]{#1}%
\providecommand \href@noop [0]{\@secondoftwo}%
\providecommand \href [0]{\begingroup \@sanitize@url \@href}%
\providecommand \@href[1]{\@@startlink{#1}\@@href}%
\providecommand \@@href[1]{\endgroup#1\@@endlink}%
\providecommand \@sanitize@url [0]{\catcode `\\12\catcode `\$12\catcode
  `\&12\catcode `\#12\catcode `\^12\catcode `\_12\catcode `\%12\relax}%
\providecommand \@@startlink[1]{}%
\providecommand \@@endlink[0]{}%
\providecommand \url  [0]{\begingroup\@sanitize@url \@url }%
\providecommand \@url [1]{\endgroup\@href {#1}{\urlprefix }}%
\providecommand \urlprefix  [0]{URL }%
\providecommand \Eprint [0]{\href }%
\@ifxundefined \urlstyle {%
  \providecommand \doi  [0]{\begingroup \@sanitize@url \@doi}%
  \providecommand \@doi [1]{\endgroup \@@startlink {\doibase
  #1}doi:\discretionary {}{}{}#1\@@endlink }%
}{%
  \providecommand \doi  [0]{doi:\discretionary{}{}{}\begingroup
  \urlstyle{rm}\Url }%
}%
\providecommand \doibase [0]{http://dx.doi.org/}%
\providecommand \Doi [0]{\begingroup \@sanitize@url \@Doi }%
\providecommand \@Doi  [1]{\endgroup\@@startlink{\doibase#1}\@@Doi}%
\providecommand \@@Doi [1]{#1\@@endlink}%
\providecommand \selectlanguage [0]{\@gobble}%
\providecommand \bibinfo  [0]{\@secondoftwo}%
\providecommand \bibfield  [0]{\@secondoftwo}%
\providecommand \translation [1]{[#1]}%
\providecommand \BibitemOpen [0]{}%
\providecommand \bibitemStop [0]{}%
\providecommand \bibitemNoStop [0]{.\EOS\space}%
\providecommand \EOS [0]{\spacefactor3000\relax}%
\providecommand \BibitemShut  [1]{\csname bibitem#1\endcsname}%
\bibitem [{\citenamefont {Smith}\ \emph {et~al.}(2005)\citenamefont {Smith},
  \citenamefont {Mock}, \citenamefont {Starr},\ and\ \citenamefont
  {Schurig}}]{Smith2005Gradient}%
  \BibitemOpen
  \bibfield  {author} {\bibinfo {author} {\bibfnamefont {D.~R.}\ \bibnamefont
  {Smith}}, \bibinfo {author} {\bibfnamefont {J.~J.}\ \bibnamefont {Mock}},
  \bibinfo {author} {\bibfnamefont {A.~F.}\ \bibnamefont {Starr}}, \ and\
  \bibinfo {author} {\bibfnamefont {D.}~\bibnamefont {Schurig}},\ }\bibfield
  {title} {\enquote {\bibinfo {title} {Gradient index metamaterials},}\
  }\href@noop {} {\bibfield  {journal} {\bibinfo  {journal} {Phys. Rev. E},\
  }\textbf {\bibinfo {volume} {71}},\ \bibinfo {pages} {036609} (\bibinfo
  {year} {2005})}\BibitemShut {NoStop}%
\bibitem [{\citenamefont {Yu}\ \emph {et~al.}(2011)\citenamefont {Yu},
  \citenamefont {Genevet}, \citenamefont {Kats}, \citenamefont {Aieta},
  \citenamefont {Tetienne}, \citenamefont {Capasso},\ and\ \citenamefont
  {Gaburro}}]{Yu2011Light}%
  \BibitemOpen
  \bibfield  {author} {\bibinfo {author} {\bibfnamefont {N.}~\bibnamefont
  {Yu}}, \bibinfo {author} {\bibfnamefont {P.}~\bibnamefont {Genevet}},
  \bibinfo {author} {\bibfnamefont {M.~A.}\ \bibnamefont {Kats}}, \bibinfo
  {author} {\bibfnamefont {F.}~\bibnamefont {Aieta}}, \bibinfo {author}
  {\bibfnamefont {J.-P.}\ \bibnamefont {Tetienne}}, \bibinfo {author}
  {\bibfnamefont {F.}~\bibnamefont {Capasso}}, \ and\ \bibinfo {author}
  {\bibfnamefont {Z.}~\bibnamefont {Gaburro}},\ }\bibfield  {title} {\enquote
  {\bibinfo {title} {Light propagation with phase discontinuities: Generalized
  laws of reflection and refraction},}\ }\href@noop {} {\bibfield  {journal}
  {\bibinfo  {journal} {Science},\ }\textbf {\bibinfo {volume} {334}},\
  \bibinfo {pages} {333} (\bibinfo {year} {2011})}\BibitemShut {NoStop}%
\bibitem [{\citenamefont {Ni}\ \emph {et~al.}(2012)\citenamefont {Ni},
  \citenamefont {Emani}, \citenamefont {Kildishev}, \citenamefont
  {Boltasseva},\ and\ \citenamefont {Shalaev}}]{Ni2012Broadband}%
  \BibitemOpen
  \bibfield  {author} {\bibinfo {author} {\bibfnamefont {X.}~\bibnamefont
  {Ni}}, \bibinfo {author} {\bibfnamefont {N.~K.}\ \bibnamefont {Emani}},
  \bibinfo {author} {\bibfnamefont {A.~V.}\ \bibnamefont {Kildishev}}, \bibinfo
  {author} {\bibfnamefont {A.}~\bibnamefont {Boltasseva}}, \ and\ \bibinfo
  {author} {\bibfnamefont {V.~M.}\ \bibnamefont {Shalaev}},\ }\bibfield
  {title} {\enquote {\bibinfo {title} {Broadband light bending plasmonic
  nanoantennas},}\ }\href@noop {} {\bibfield  {journal} {\bibinfo  {journal}
  {Science},\ }\textbf {\bibinfo {volume} {335}},\ \bibinfo {pages} {427}
  (\bibinfo {year} {2012})}\BibitemShut {NoStop}%
\bibitem [{\citenamefont {Magnusson}\ and\ \citenamefont
  {Gaylord}(1978)}]{Magnusson1978Diffraction}%
  \BibitemOpen
  \bibfield  {author} {\bibinfo {author} {\bibfnamefont {R.}~\bibnamefont
  {Magnusson}}\ and\ \bibinfo {author} {\bibfnamefont {T.~K.}\ \bibnamefont
  {Gaylord}},\ }\bibfield  {title} {\enquote {\bibinfo {title} {Diffraction
  efficiencies of thin phase gratings with arbitrary grating shape},}\
  }\href@noop {} {\bibfield  {journal} {\bibinfo  {journal} {J. Opt. Soc.
  Am.},\ }\textbf {\bibinfo {volume} {68}},\ \bibinfo {pages} {806} (\bibinfo
  {year} {1978})}\BibitemShut {NoStop}%
\bibitem [{\citenamefont {Lalanne}\ \emph {et~al.}(1998)\citenamefont
  {Lalanne}, \citenamefont {Astilean}, \citenamefont {Chavel}, \citenamefont
  {Cambril},\ and\ \citenamefont {Launois}}]{Lalanne1998Blazed}%
  \BibitemOpen
  \bibfield  {author} {\bibinfo {author} {\bibfnamefont {P.}~\bibnamefont
  {Lalanne}}, \bibinfo {author} {\bibfnamefont {S.}~\bibnamefont {Astilean}},
  \bibinfo {author} {\bibfnamefont {P.}~\bibnamefont {Chavel}}, \bibinfo
  {author} {\bibfnamefont {E.}~\bibnamefont {Cambril}}, \ and\ \bibinfo
  {author} {\bibfnamefont {H.}~\bibnamefont {Launois}},\ }\bibfield  {title}
  {\enquote {\bibinfo {title} {Blazed binary subwavelength gratings with
  efficiencies larger than those of conventional \'echelette gratings},}\
  }\href@noop {} {\bibfield  {journal} {\bibinfo  {journal} {Opt. Lett.},\
  }\textbf {\bibinfo {volume} {23}},\ \bibinfo {pages} {1081} (\bibinfo {year}
  {1998})}\BibitemShut {NoStop}%
\bibitem [{\citenamefont {Tsai}\ \emph {et~al.}(2011)\citenamefont {Tsai},
  \citenamefont {Larouche}, \citenamefont {Tyler}, \citenamefont {Lipworth},
  \citenamefont {Jokerst},\ and\ \citenamefont {Smith}}]{Tsai2011Design}%
  \BibitemOpen
  \bibfield  {author} {\bibinfo {author} {\bibfnamefont {Y.-J.}\ \bibnamefont
  {Tsai}}, \bibinfo {author} {\bibfnamefont {S.}~\bibnamefont {Larouche}},
  \bibinfo {author} {\bibfnamefont {T.}~\bibnamefont {Tyler}}, \bibinfo
  {author} {\bibfnamefont {G.}~\bibnamefont {Lipworth}}, \bibinfo {author}
  {\bibfnamefont {N.~M.}\ \bibnamefont {Jokerst}}, \ and\ \bibinfo {author}
  {\bibfnamefont {D.~R.}\ \bibnamefont {Smith}},\ }\bibfield  {title} {\enquote
  {\bibinfo {title} {Design and fabrication of a metamaterial gradient index
  diffraction grating at infrared wavelengths},}\ }\href@noop {} {\bibfield
  {journal} {\bibinfo  {journal} {Opt. Express},\ }\textbf {\bibinfo {volume}
  {19}},\ \bibinfo {pages} {24411} (\bibinfo {year} {2011})}\BibitemShut
  {NoStop}%
\bibitem [{\citenamefont {Pendry}(2000)}]{Pendry2000Negative}%
  \BibitemOpen
  \bibfield  {author} {\bibinfo {author} {\bibfnamefont {J.~B.}\ \bibnamefont
  {Pendry}},\ }\bibfield  {title} {\enquote {\bibinfo {title} {Negative
  refraction makes a perfect lens},}\ }\href@noop {} {\bibfield  {journal}
  {\bibinfo  {journal} {Phys. Rev. Lett.},\ }\textbf {\bibinfo {volume} {85}},\
  \bibinfo {pages} {3966} (\bibinfo {year} {2000})}\BibitemShut {NoStop}%
\bibitem [{\citenamefont {Smith}\ \emph {et~al.}(2011)\citenamefont {Smith},
  \citenamefont {Tsai},\ and\ \citenamefont {Larouche}}]{Smith2011Analysis}%
  \BibitemOpen
  \bibfield  {author} {\bibinfo {author} {\bibfnamefont {D.~R.}\ \bibnamefont
  {Smith}}, \bibinfo {author} {\bibfnamefont {Y.-J.}\ \bibnamefont {Tsai}}, \
  and\ \bibinfo {author} {\bibfnamefont {S.}~\bibnamefont {Larouche}},\
  }\bibfield  {title} {\enquote {\bibinfo {title} {Analysis of a gradient index
  metamaterial blazed diffraction grating},}\ }\href@noop {} {\bibfield
  {journal} {\bibinfo  {journal} {IEEE Antennas Wireless Prop. Lett.},\
  }\textbf {\bibinfo {volume} {10}},\ \bibinfo {pages} {1605} (\bibinfo {year}
  {2011})}\BibitemShut {NoStop}%
\end{thebibliography}%

\end{document}